\DocumentMetadata{lang={en-GB},tagging=on}
\documentclass[a4paper]{article}
\usepackage[T1,T2A]{fontenc}
\usepackage[
 unicode,
 pdfauthor={David Simon Henrik Jonsson, 金炯錄},
 pdftitle={Hydrodynamics as cospans of field theories into the 𝐵𝐹 theory},
 pdfsubject={{hep-th, cond-mat.str-el, math-ph}},
 breaklinks=true,
 final
 ]{hyperref}
\usepackage{amsmath,amssymb,orcidlink,cleveref,tikz-cd,mleftright,mathtools}
\usepackage[noadjust]{cite}
\usetikzlibrary{babel}
\usepackage{CJKutf8}
\usepackage[final]{microtype}
\usepackage[latin,russian,french,german,vietnamese,british]{babel}
\usepackage[osf]{newpxtext}
\usepackage[varbb,slantedGreek]{newpxmath}
\usepackage[utf8]{inputenc}

\renewcommand\pi\piup
\renewcommand\delta\deltaup
\renewcommand\Omega\Omegaup
\renewcommand\Gamma\Gammaup
\renewcommand\varepsilon\varepsilonup
\newcommand\intprod{\mathbin\lrcorner}
\newcommand\Ep[1]{E^{(#1)}}

\begin{document}
\title{Hydrodynamics as cospans of field theories into the \(BF\) theory}
\author[pdfauthor={{David Simon Henrik Jonsson, 金炯錄}}]
{David Simon Henrik Jonsson\textsuperscript{\orcidlink{0009-0001-7155-8496}}\footnote{Centre for Mathematics and Theoretical Physics Research, Department of Physics, Astronomy and Mathematics, University of Hertfordshire, Hatfield, Hertfordshire\ \textsc{al10 9ab}, United Kingdom}\\
\href{simon.jonsson010@gmail.com}{\texttt{simon.jonsson010@gmail.com}}\\\\
Hyungrok Kim~(\begin{CJK*}{UTF8}{bsmi}金炯錄\end{CJK*})\textsuperscript{\orcidlink{0000-0001-7909-4510}}\footnotemark[1]\\
\href{mailto:h.kim2@herts.ac.uk}{\texttt{h.kim2@herts.ac.uk}}
}
\maketitle
\begin{abstract}
Hydrodynamics is based on conservation laws of currents: one starts from the conserved currents of the theory describing the microscopic dynamics, and provides an alternative parameterisation of these currents in terms of hydrodynamic variables (density, pressure, velocity, etc.). This paradigm has recently been extended to incorporate higher-form symmetries \cite{Grozdanov:2016tdf,Armas:2018ibg,Armas:2018atq,Delacretaz:2019brr}. The conservation law of the \(p\)-form conserved currents can be regarded as the equations of motion of a \(BF\) theory that treats the currents as fundamental fields. We argue that the hydrodynamic approximation to a microscopic theory can be regarded as a cospan of differential graded manifolds \(X_\mathrm{micro}\to X_{BF}\leftarrow X_\mathrm{hydro}\), where \(X_\mathrm{micro}\) and \(X_\mathrm{hydro}\) describe the microscopic and hydrodynamic theories, respectively, and \(X_{BF}\) describes the \(BF\) theory of conserved currents.
\end{abstract}
\tableofcontents

\section{Introduction and summary}\label{sec:intro}
Hydrodynamics \cite{landau,arnold} provides a universal description of many-body systems at long times and wavelengths. Apart from the enormously successful theory of nonrelativistic hydrodynamics, relativistic hydrodynamics \cite{Gourgoulhon:2006bn,Font:2007zza} has been used to successfully describe field-theoretic systems in heavy-ion collisions \cite{Ollitrault:2007du,Hirano:2008hy}, astrophysics \cite{Andersson:2006nr}, and beyond. Today, hydrodynamics is understood as a consequence of conservation laws associated to symmetries. Furthermore, the notion of `symmetry' has been considerably generalised in recent years \cite{Gaiotto:2014kfa} (see the reviews \cite{Bhardwaj:2023kri,Gomes:2023ahz,Cordova:2022ruw,Brennan:2023mmt,Luo:2023ive,Iqbal:2024pee}), and such generalised symmetries play a key role in the hydrodynamic description of many systems, including magnetohydrodynamics \cite{Grozdanov:2016tdf,Armas:2018ibg,Armas:2018atq,Delacretaz:2019brr}.

The microscopic degrees of freedom, on the other hand, should be described by a quantum field theory, such as quantum chromodynamics, typically with gauge symmetries that can be naturally described by the Batalin--Vilkovisky formalism \cite{Batalin:1977pb,Batalin:1983ggl,Batalin:1984ss,Batalin:1985qj} (reviewed in \cite{Henneaux:1994lbw,Henneaux:1989jq,Gomis:1994he,Barnich:2000zw,Fiorenza:2004sg,Fuster:2005eg,Mnev:2017oko,Barnich:2018gdh}); in particular, the Keldysh--Schwinger formalism, a key ingredient in microscopic derivations of relativistic hydrodynamics, naturally arises in the Batalin--Vilkovisky formalism \cite{Borsten:2024dvq}. It is therefore a natural question to seek a formalisation of the matching of the microscopic and macroscopic descriptions of the same physics, with the microscopic theory on one side and the hydrodynamic theory on the other. This paper describes this matching as a cospan of differential graded manifolds:
\begin{equation}
        \begin{tikzcd}[row sep = 1 ex, ampersand replacement=\&]
          \label{eq:cospan}
          \& \text{\(BF\) theory} \&  \\
                \text{microscopic theory} \ar{ur} \& \&\ar{ul} \text{hydrodynamic theory}
        \end{tikzcd},
\end{equation}
where the object in the middle is an Abelian \(BF\) theory \cite{Broda:2005wk}, a topological field theory whose equations of motion are simply that \(p\)-form fields --- corresponding to \(p\)-form Noether currents of continuous generalised symmetries--- are closed.
As a simple example, on a \(d\)-dimensional pseudo-Riemannian manifold \((M,g)\), one considers the conservation of the stress--energy tensor:
\begin{equation}
    \nabla_\mu T^{\mu\nu}=0.
\end{equation}
This is a set of \(d\) differential equations with respect to the \(d(d+1)/2\) components of \(T^{\mu\nu}\); to obtain a closed set of equations, one parameterises \(T^{\mu\nu}\) in terms of \(d\) variables --- for instance, we can consider the perfect-fluid parameterisation \cite[§5.5]{Misner:1973prb}
\begin{equation}\label{eq:perfect-fluid}
    T^{\mu\nu} = (\rho + p(\rho))u^\mu u^\nu + p(\rho) g^{\mu\nu}
\end{equation}
where \(p(\rho)\) is a function of the density \(\rho\) (the (function appearing in the) equation of state\footnote{Following common abuse of terminology, we refer to the function \(p(\rho)\) as the `equation of state'.}), and \(u^\mu\) (the four-velocity field of the fluid) is normalised to have unit norm; then \((\rho,u^\mu)\) provides a set of \(d\) independent variables, so that
\begin{equation}
    \nabla_\mu T^{\mu\nu}(\rho,u) = 0
\end{equation}
is a closed set of equations.\footnote{One can generalise \eqref{eq:perfect-fluid} to include derivative terms of \(u^\mu\) and in so doing include viscosity and transport coefficients.}

Furthermore, one can generalise the above system by including other conservation laws, such as electric charge. For a charged fluid, one then has the \(d+1\) equations
\begin{align}\label{eq:charged-fluid-example}
    \nabla_\mu T^{\mu\nu}&=0,&\nabla_\mu J^\mu&=0
\end{align}
with \(T^{\mu\nu}\) and \(J^\mu\) parameterised by a set of \(d+1\) independent parameters, such as \((\rho,u^mu,\rho_{\mathrm e})\), where \(\rho_{\mathrm e}\) is the charge density. Note that assuming that \(M\) is oriented, we can use the language of differential forms.
If we Hodge-dualise one Lorentz index from \(T^{\mu\nu}\) and \(J^\mu\), then \eqref{eq:charged-fluid-example} can be recast as 
\begin{align}
  \label{eq:0-form}
    \mathrm dT^\nu &= 0, & \mathrm dJ&=0,
\end{align}
where \(T^\nu\) and \(J\) are (vector-valued, in the case of \(T^\mu\)) differential \((d-1)\)-forms.

This formulation naturally lends itself to generalisation. In particular, it extends straightforwardly to the case of higher-form symmetries \cite{Gaiotto:2014kfa} (see the reviews \cite{Bhardwaj:2023kri,Gomes:2023ahz,Cordova:2022ruw,Brennan:2023mmt,Luo:2023ive,Iqbal:2024pee}).

In the special case of continuous (rather than discrete) symmetries, which is the case this paper is concerned with, this generalisation entails allowing Noether currents that are not \((d-1)\)-forms but rather \(p\)-forms of arbitrary form degree \(p\). (The somewhat confusing but standard terminology is that a `\(k\)-form symmetry' corresponds to a Noether current that is a closed \(p\)-form where \(p=(d-k-1)\).)
Suppose that we consider a microscopic quantum field theory that enjoy a set of higher-form symmetries whose conserved Noether currents are \(\{J^{(p)}\}_{p=0}^{d-1}\) of \(p\)-forms \(J^{(p)}\),
valued in some \(N_p\)-dimensional vector bundles \(E^{(p)}\),\footnote{As an exception, \(E^{(0)}\) can be a smooth fibre bundle that is not necessarily a vector bundle.} where \(J^{(p)}\) is the Noether current of a \((d-p-1)\)-form symmetry.

The conservation laws associated to the symmetries state that these Noether currents are closed:\footnote{
    As discussed further in \cref{ssec:BF-model}, when the bundles \(E^{(p)}\) are not flat, in general \eqref{eq:closed-current} requires the choice of a connection on \(E^{(p)}\).
}
\begin{equation}\label{eq:closed-current}
    \mathrm dJ^{(p)} = 0.
\end{equation}
This constitutes a set of \(N\) differential equations, where
\begin{equation}
    N = \sum_p N_p\binom d{p+1},
\end{equation}
and one parameterises the currents \((J^{(0)},\dotsc,J^{(d-1)})\) using \(N\) independent variables.

The equations \eqref{eq:closed-current} can be derived from an action principle
\begin{equation}\label{eq:BF-action}
    S_{BF} = \int_M \Lambda^{(d-p-1)}\wedge \mathrm dJ^{(p)},
\end{equation}
whose fundamental fields are the \(p\)-form currents \(J^{(p)}\) and corresponding \((d-p-1)\)-form Lagrange multipliers \(\Lambda^{(d-p-1)}\). The action \eqref{eq:BF-action} is in the form of a \(BF\) theory (reviewed in \cite{Broda:2005wk}), except that there are multiple pairs of gauge fields \(J^{(p)}\) and \(\Lambda^{(d-p-1)}\) and that they take values in nontrivial bundles if the conserved currents are valued in nontrivial bundles.\footnote{
    In ordinary \(BF\) theory, the fields are differential forms valued in the trivial real line bundle and which enjoy gauge symmetries.
    In our case, if the bundle \(E^{(p)}\) does not have a flat connection, then in general the action \eqref{eq:BF-action} does not have gauge symmetries, as explained in \cref{ssec:BF-model}.
}

In the \(BF\) action \eqref{eq:BF-action}, the conserved currents \(J^{(p)}\) appear as fundamental fields, severed from the microscopic theory. Of course, from the perspective of the microscopic theory, the conserved currents \(J^{(p)}\) are not fundamental but rather functionals \(J^{(p)}[\phi]\) of the microscopic fields, which we label as \(\phi\). This correspondence between the fields \(\phi\) of the microscopic theory and the fields \(J^{(p)}\) of the \(BF\) theory gives rise to a map between theories
\begin{equation}
    \text{microscopic theory}\to \text{\(BF\) theory},
\end{equation}
which we formalise as a morphism of differential graded manifolds.

Similarly, hydrodynamics can then be described by the choice\footnote{Not every possible choice of parameterisation is equally good; some choices may be more preferable than others on physical or mathematical grounds. We do not comment on this issue further here.} of a parameterisation of the conserved currents \(J^{(p)}\) using hydrodynamic variables \(\rho, u^i,\dotsc\), so that one can regard \(J^{(p)}[\rho, u^i,\dotsc]\) as a functional of the hydrodynamic variables. The conservation law
\begin{equation}\label{eq:hydrodynamic-conservation-law}
    \mathrm dJ^{(p)}[\rho,u^i,\dotsc]=0
\end{equation}
holds in both the \(BF\) theory (where it is an Euler--Lagrange equation derived from the action \eqref{eq:BF-action}) and in the hydrodynamic theory (where the equation \eqref{eq:hydrodynamic-conservation-law} is directly imposed, without reference to an action); this gives rise to a map
\begin{equation}
    \text{hydrodynamic theory}\to \text{\(BF\) theory}.
\end{equation}

The above discussion gives rise to the following diagram (or \emph{cospan}, in category-theoretic language; see \cite[Exercise~2.7.7]{johnson20202dimensionalcategories} or \cite[p.~284]{maclane}) of field theories:
  \begin{equation}\label{eq:informal-cospan}
        \begin{tikzcd}[row sep = 1 ex, ampersand replacement=\&]
          \& \text{\(BF\) theory} \&  \\
                \text{microscopic theory} \ar{ur} \& \&\ar{ul} \text{hydrodynamic theory},
        \end{tikzcd}
      \end{equation}
where
\begin{itemize}
\item the `microscopic theory' is the possibly complicated quantum field theory

(e.g.\ quantum electrodynamics, Yang--Mills, …) describing the microscopic degrees of freedom with fields \(\phi\);\footnote{In order to deal with the quantum effects, one should either introduce the Batalin--Vilkovisky Laplacian and solve the quantum master equation \cite{Henneaux:1994lbw}, or instead work with a one-particle-irreducible effective action \cite{Borsten:2025hrn}.}
\item the morphism `\(\text{microscopic theory} \to \text{\(BF\) theory}\)' maps the fields \(\phi\) of the microscopic theory to the currents \(J[\phi]\) that appear in the \(BF\) theory;
\item the morphism `\(\text{hydrodynamic theory} \to \text{\(BF\) theory}\)' maps the variables \(\rho,u^\mu,\dotsc\) of the hydrodynamic theory to the currents \(J[\rho,u^\mu,\dotsc]\) that appear in the \(BF\) theory.
\end{itemize}
This cospan can then be elegantly formulated in the language of differential graded geometry and the Batalin--Vilkovisky formalism, which is the main result of this paper.

\paragraph{Limitations and future directions.}
In this paper, we have neglected thermodynamic considerations. That is, the conservation equations we deal with do not include the conservation of entropy. While hydrodynamic conservation laws can be readily written for the conservation of specific entropy, the morphism \(\text{microscopic theory} \to \text{\(BF\) theory}\) would then need to involve the formalism of thermal field theory to define entropy for the microscopic theory, which we do not delve into for this paper.

This paper concerns itself with solely those parts of hydrodynamics that can be reduced to conservation laws. Techniques from holography \cite{Ambrosetti:2008mt,Hubeny:2010wp,Hubeny:2011hd} provide insights on various transport coefficients including a universal lower bound on viscosity \cite{Kovtun:2004de} (see the review \cite{Schafer:2009dj}), which are outside the scope of this paper.

From the perspective of generalised symmetries, we have conspicuously restricted ourselves to continuous, invertible symmetries, rather than discrete symmetries that appear in many systems. The discrete nature of these symmetries means that they do not have ordinary differential-form-valued Noether currents, so that the techniques of differential equations no longer apply. Nevertheless, at least for discrete invertible symmetries, there is a \(BF\) description of discrete-valued generalised symmetries using cohomology theories with discrete rather than continuous coefficients, viz.\ we can use simplicial or Čech cocycles rather than de~Rham cocycles (i.e.\ closed differential forms). However, it is not immediate if this can lead to a `hydrodynamic' description.

\paragraph{Organisation of this paper.} After a brief summary that avoids the language of differential graded geometry in \cref{sec:intro}, we quickly review the relevant geometric concepts and fix conventions in \cref{sec:review}, using which we formalise hydrodynamics as a cospan of differential graded manifolds in \cref{sec:cospan}, including a uniform treatment of the higher-form case. In \cref{sec:examples}, we connect our formalism to ordinary formulations of hydrodynamics by recasting well known hydrodynamic equations into our language.

\paragraph{Conventions and notation.} We use the mostly plus sign convention for the pseudo-Riemannian metric \(g_{\mu\nu}=\operatorname{diag}(-1,1,\dotsc,1)\). The notation \(\intprod\) denotes the interior product between a vector field and a differential form. Given a vector field \(v^\mu\), the notation \(v^\flat\) denotes the corresponding differential one-form \(v^\mu g_{\mu\nu}\mathrm dx^\nu\) using the musical isomorphism given by a (pseudo-)Riemannian metric \(g_{\mu\nu}\). Symmetrisation of indices is normalised: \(X_{(\mu\nu)}=\frac12(X_{\mu\nu}+X_{\nu\mu})\) etc.

For graded objects, we adhere to the Koszul sign rule \cite{Deligne:1999qp}. Given a graded object \(V=\bigoplus_iV_i\), The notation \(V[k]\) denotes suspension, such that \((V[k])_i = V_{k+i}\). Thus, if \(V\) is an ungraded object regarded as being concentrated in degree zero, then \(V[k]\) is concentrated in degree \(-k\). Space of differential forms \(\Omega^k(M)\) do \emph{not} carry inherent degree; thus \(\Omega^k(M)[-1]\) is concentrated in degree \(1\), not in \(k+1\).

\section{Preliminaries}\label{sec:review}
The Batalin--Vilkovisky formalism offers a unified framework for quantum field theories, including those with gauge symmetries, and is formulated in the language of differential graded geometry. We provide a brief informal review below and refer the reader to \cite{Henneaux:1994lbw,Henneaux:1989jq,Gomis:1994he,Barnich:2000zw,Fiorenza:2004sg,Fuster:2005eg,Mnev:2017oko,Barnich:2018gdh} for more detailed treatments.

\subsection{Differential graded geometry}
Differential graded geometry \cite{Cattaneo:2010re,Qiu:2011qr} is a generalisation of ordinary differential geometry and supergeometry in which coordinates can carry degree (such that odd coordinates anticommute amongst themselves), and a choice of a \emph{homological vector field} provides a notion of what is physical or unphysical similar to BRST cohomology.

An ordinary \(d\)-dimensional manifold \(M\) is modelled on subsets of \(\mathbb R^d\), i.e.\ can be thought of as glued together from a family of open subsets \(U_i\subset\mathbb R^d\); the components \((x^1,\dotsc,x^d)\) of the partially defined map \(M\supset U_i \to \mathbb R^d\) then serve as local coordinates on \(M\).
Similarly, a \emph{graded manifold} \(M\) is modelled on graded vector spaces, i.e.\ glued together from a family of patches
\begin{equation}
  V_i=U_i\times \bigoplus_{n_i\in\mathbb Z}\mathbb R^{n_i}[-i].
\end{equation}
Its local coordinates \((x^1,x^2,\dotsc)\colon M\to V_i\) therefore carry integer degrees \(|x_i|\in\mathbb Z\) such that coordinates with odd degrees are Grassmann-valued (i.e.\ anticommute with each other) while coordinates with even degrees are commuting (i.e.\ commute with each other and with odd-degree coordinates).

A \emph{differential graded manifold}, or \emph{dg-manifold} for short, is a graded manifold \(X\) equipped with a vector field, called the \emph{homological vector field},
\begin{equation}Q = Q^i\frac\partial{\partial x^i}\end{equation}
of degree \(|Q|=1\) such that its Lie bracket with itself vanishes:
\begin{equation}[Q,Q]=0.\end{equation}

A \emph{symplectic differential graded manifold} is a differential graded manifold \((X,Q)\) equipped with a symplectic structure \(\omega\) of degree \(-1\) that pairs fields with their corresponding antifields.\footnote{
On noncompact spacetimes, there are analytic issues in making \(\omega\) well defined for arbitrary fields and antifields with noncompact support; see e.g.\ \cite{Macrelli:2019afx}. We largely gloss over these issues here.} Then the notion of a differential graded symplectomorphism between differential graded symplectic manifolds is defined the obvious way, viz.\ as a differential graded map \(f\colon (X,Q_X,\omega_X)\to(Y,Q_Y,\omega_Y)\) such that the pullback of the symplectic form on \(Y\) agrees with that on \(X\), that is, \(f^*\omega_Y=\omega_X\).

The tangent space of a differential graded manifold admits the structure of a (curved) \(L_\infty\)-algebra \cite{Jurco:2018sby,Kraft:2022efy}.
When the differential graded manifold further carries a symplectic structure, this corresponds to a cyclic structure on the \(L_\infty\)-algebra.

\subsection{The Batalin--Vilkovisky formalism}\label{ssec:BV}
In the Batalin--Vilkovisky formalism, in order to consistently treat gauge symmetries, the collection of physical fields is extended in two directions: first, corresponding to gauge symmetries, one introduces Faddeev--Popov ghost fields (and higher-order ghost fields, if the ghost fields also gauge-transform).
This defines a graded manifold \(X_\mathrm{BRST}\), whose coordinates are physical fields and Faddeev--Popov ghosts, equipped with a vector field \(Q_\mathrm{BRST}\), the BRST differential. Generally, \(Q_\mathrm{BRST}\) squares to zero only up to the equations of motion, so that \(X_\mathrm{BRST}\) is not generally a differential graded manifold.

The Batalin--Vilkovisky formalism remedies this by the introduction of \emph{antifields}. Formally, one considers the (fibrewise) degree-shifted cotangent bundle of \(X_\mathrm{BRST}\):
\begin{equation}
  X\coloneq\mathrm T^*[-1]X_\mathrm{BRST}.
\end{equation}
Then \(X\) naturally carries a symplectic form \(\omega\) of degree \(-1\) that pairs a physical field or Faddeev--Popov ghost field \(\phi\) with its corresponding antifield \(\phi^+\). The BRST differential $Q_\mathrm{BRST}$ is then extended to a differential $Q$ on $X$ such that \(Q\) applied to a functional \(f(\phi)\) of the fields (of degree zero) yields an object valued in the antifields since the equations of motion are in canonical bijection with the antifields. 
The collection of physical fields, ghost fields, and their corresponding antifields then forms an infinite-dimensional differential graded symplectic manifold, where the grading corresponds to the ghost numbers of the fields and antifields. In our conventions, the coordinate degree \(|x^i|\) is the negative of the usual convention for the ghost number: degree-zero coordinates correspond to ordinary fields; degree-one coordinates corresponds to antifields; degree-\((-1)\) coordinates correspond to Faddeev--Popov ghosts; coordinates of degrees \(n\le-2\) correspond to \((-n)\)th-order ghosts; coordinates of degrees \(n\ge2\) correspond to antifields of \((n-1)\)st-order ghosts. So, for example, a theory of a scalar field (together with its antifield) on a smooth manifold \(M\) is described by the infinite-dimensional graded manifold
\begin{equation}
    X = \mathcal C^\infty(M;\mathbb R)\times\mathcal C^\infty(M;\mathbb R)[-1],
\end{equation}
where the former factor describes the scalar field \(\phi\) while the latter factor corresponding to its antifield \(\phi^+\) sits in degree \(1\).

A classical\footnote{
    In general, one must supplement this with \(\hbar\)-dependent counterterms to produce the quantum action that satisfies the quantum master equation; we do not discuss this issue here.
} \emph{action} \(S\) on a symplectic differential graded manifold \((X,Q,\omega)\) is a function\footnote{
    For essentially the same reasons that the symplectic structure \(\omega\) is only partially defined when spacetime is noncompact, similarly on a noncompact spacetime the action can diverge for field configurations with noncompact support and is hence, strictly speaking, not defined on \emph{all} of \(X\). For rigorous approaches using cut-off functions, see e.g.\ \cite[§4.1]{Rejzner:2016hdj} or \cite[§1.5]{Dutsch:2019wxk}.} on \(X\) of degree zero such that
\begin{align}
    (\omega^{-1})^{ij}\partial_j S&= Q^i,
\end{align}
where \(\omega^{-1}\) is the inverse of (i.e.\ the Poisson structure corresponding to) the symplectic form \(\omega\). Turning things backwards, from the choice of a classical action \(S\) and the symplectic structure \(\omega\), one can construct the corresponding homological vector field \(Q\).

The Batalin--Vilkovisky formalism, then, describes a classical field theory admitting an action principle as simply a symplectic differential graded manifold whose symplectic structure is of degree \(-1\).
More generally, those theories that are simply given by equations of motion without an action principle are instead described by differential graded manifolds; the existence of a compatible symplectic structure corresponds to the ability to formulate an action principle for the theory.

\subsection{Differential forms on non-orientable manifolds}
Hydrodynamics can be defined over any smooth manifold regardless of its orientability.
The usual definition of differential forms, however, requires a choice of orientation. In order to talk about \(p\)-form Noether currents on manifolds without an orientation, we make use of the \emph{orientation bundle} \cite[§2.9.5]{kashiwara} and differential forms twisted with respect to it.

Let \(M\) be a smooth manifold of dimension \(d\), where we do not assume an orientation.
The \emph{orientation double cover} \(\operatorname{Or}(M)\) is the double cover of \(M\) given by
\begin{equation}
    \operatorname{Or}(M)=\{(x,o)| x\in M,\, o\in \operatorname{Or}_x\}\to M,
\end{equation}
where \(\operatorname{Or}_x\) is the set of the two local orientations near \(x\).
The group of order two \(\mathbb Z_2\) acts on \(\operatorname{Or}(M)\) by reversing the local orientations; this makes \(\operatorname{Or}(M)\) a principal \(\mathbb Z_2\)-bundle over \(M\). The \emph{orientation bundle} is the real line bundle on \(M\) that is the associated bundle of \(\operatorname{Or}(M)\) with respect to the real one-dimensional representation \(\mathbb Z_2\to\{\pm1\}\):
\begin{equation}
\label{eq:orientation bundle}
    \operatorname{or}_M\coloneqq\operatorname{Or}(M)\times_{\mathbb Z_2}\mathbb R.
\end{equation}
By construction, this is a flat line bundle equipped with a canonical metric, and its tensor square \(\operatorname{or}_M\otimes_M\operatorname{or}_M=M\times\mathbb R\) is the trivial line bundle.
In particular, we can define \(\operatorname{or}_M\)-valued differential forms (sometimes called `pseudo-differential forms' or `twisted differential forms'):
\begin{equation}
    \Omega^k(M;\operatorname{or}_M)\coloneq\Gamma_M\mleft(\bigwedge^k\mathrm T^*M\otimes \operatorname{or}_M\mright).
\end{equation}

Unlike an ordinary top-degree differential form (i.e.\ differential \(d\)-form), a \(\operatorname{or}_M\)-valued differential \(d\)-form can be integrated on \(M\) without the choice of an orientation.
Furthermore, if \(M\) is equipped with a (pseudo-)Riemannian metric \(g\), then
the Hodge-star operator maps ordinary differential forms to \(\operatorname{or}_M\)-valued differential forms and vice versa.
That is, if \(\alpha=(p!)^{-1}\alpha_{\mu_1\dotso\mu_p}\,\mathrm dx^{\mu_1}\wedge\dotsm\wedge\mathrm dx^{\mu_p}\) is an ordinary differential \(p\)-form, in the expression for its Hodge dual defined with respect to a local coordinate system \((x^\mu)\), namely
\begin{equation}
    \star\alpha\coloneqq \frac{\sqrt{|\det g|}}{p!(d-p)!}\alpha_{\mu_1\dotso\mu_p}g^{\mu_1\nu_1}\dotsm g^{\mu_p\nu_p}\varepsilon_{\nu_1\dotso\nu_d}\,\mathrm dx^{\nu_{p+1}}\wedge\dotsb\wedge\mathrm dx^{\nu_d},
\end{equation}
the presence of the Levi-Civita symbol \(\varepsilon_{\mu_1\dotso\mu_d}\) (which flips sign if the orientation changes) makes \(\star\alpha\) an \(\operatorname{or}_M\)-valued form. Similarly, if \(\alpha\) is an \(\operatorname{or}_M\)-valued differential \(p\)-form, then in the same expression as above, the presence of the Levi-Civita symbol makes \(\star\alpha\) an ordinary differential form.

\section{Cospans of differential graded manifolds as hydrodynamics}\label{sec:cospan}
We now formalise the informal discussion of \cref{sec:intro} in the language of differential graded geometry and the Batalin--Vilkovisky formalism, so that we obtain a cospan of differential graded manifolds
\begin{equation}
    \begin{tikzcd}[row sep = 1 ex, ampersand replacement=\&]
      \& X_{BF} \&  \\
            X_\mathrm{micro} \ar{ur} \& \&\ar{ul} X_\mathrm{hydro}
    \end{tikzcd},
\end{equation}
that formalise \eqref{eq:informal-cospan}. (Note that these arrows are \emph{not} symplectomorphisms in general: they do not preserve the relevant symplectic structures or actions, but only the differential graded structure, i.e.\ the equations of motion.)
Let \(M\) be spacetime equipped with a pseudo-Riemannian metric \(g_{\mu\nu}\). We do not assume that \(M\) is equipped with an orientation; let \(\operatorname{or}_M\) be its orientation bundle.

Note that such cospans arise in many contexts; in particular, when the differential graded manifolds may be linearised into \(L_\infty\)-algebras, cospans of differential graded manifolds dualise into spans\footnote{Cospans have become spans because the functor from (pointed) differential graded manifolds to \(L_\infty\)-algebras is contravariant, in accordance with the general principle that functors between geometry and algebra are contravariant, as seen in algebraic geometry \cite{hartshorne} or noncommutative differential geometry \cite{Gracia-Bondia:2001upu}.} of \(L_\infty\)-algebras, which are discussed in \cite{JalaliFarahani:2023sfq}. Furthermore, cospans may be thought of as a categorical generalisation of cobordisms (a cobordism between two manifolds defines a cospan in the category of manifolds); in particular, the symmetry topological field theories (SymTFTs) \cite{Apruzzi:2021nmk} that have been employed to describe generalised symmetries can be seen as cospans of differential grade manifolds between a \(d\)-dimensional ordinary field theory, a \(d\)-dimensional trivial theory of topological boundary conditions, and the \((d+1)\)-dimensional SymTFT \cite{Pulmann:2019vrw,Borsten:2025pvq}.

\subsection{The map from the microscopic theory to the \texorpdfstring{\(BF\)}{𝐵𝐹} theory}\label{ssec:BF-model}

Let us consider a quantum field theory on a $d$-dimensional manifold $M$,
described by the Batalin--Vilkovisky formalism as a symplectic differential graded manifold \((X_\mathrm{micro},Q_\mathrm{micro},\omega_\mathrm{micro})\).
Suppose that this field theory has a set of continuous generalised symmetries, whose corresponding set of Noether currents are \(\mathcal{J}=\{J^{(p)}\}_{p=0}^{d-1}\), where \(J^{(p)}\in\Omega^p(M;E^{(p)})\) is a differential \(p\)-form current (corresponding to a \((d-p-1)\)-form symmetry) valued in a graded fibre bundle\footnote{that is, a fibre bundle whose generic fibre carries the structure of a graded manifold} \(E^{(p)}\), where we assume that \(E^{(p)}\) is in fact a graded vector bundle\footnote{that is, a vector bundle whose fibres carry the structure of a graded vector space; equivalently, the direct sum \(E^{(p)}=\bigoplus_iE^{(p)}_i[-i]\) where each \(E^{(p)}_i\) is an ordinary vector bundle} for \(p\ne0\).\footnote{The grading of the bundles corresponds to allowing the conserved currents to carry nonzero ghost number, following \cite{Borsten:2025pbx}.
In the Batalin--Vilkovisky formalism, it is often unnatural and unnecessary to restrict objects to have zero ghost number.
Of course, in ordinary quantum systems, objects with nonzero ghost numbers cannot be measured experimentally;
however, in systems where the ghost number has been redefined by a topological or holomorphic twist (see e.g.\ \cite{Elliott:2020ecf}), objects with nonzero ghost numbers can be physical, as discussed in \cite{Borsten:2025pbx}.}
(The case \(p=0\) is special because a zero-form is nothing more than a function (or section), and a function can be valued in any fibre bundle, unlike differential forms of higher degree, which must be valued in a vector bundle.)
The Noether currents are thus elements of
\begin{equation}\label{eq:noethercurrents}
    \Gamma_M\mleft(\bigoplus_{p=0}^{d-1}\mathrm T^{*\wedge p}M\otimes E^{(p)}\mright)
    =
    \Omega^0(M;\Ep{0})\times\Omega^1(M;\Ep{1})\times\dotsb\times\Omega^{d-1}(M;\Ep{d-1}).
\end{equation}

For example, consider a theory where only the stress--energy tensor $T^{\mu\nu}$ is conserved. We can dualise the stress--energy tensor as
\begin{equation}\label{eq:dualise}
    T^\nu\coloneqq \star(T_\mu{}^\nu\,\mathrm dx^\mu)\in \Omega^{d-1}(M;\mathrm TM\otimes\operatorname{or}_M),
\end{equation}
which then becomes a \((d-1)\)-form valued in \(\mathrm TM\otimes\operatorname{or}_M\).
In our language we then have \(E^{(d-1)}=\mathrm TM\otimes\operatorname{or}_M\) and \(E^{(p)}\) is trivial for \(p\ne d-1\).

Now, we need to assert that the currents $J^{(p)}$ are indeed conserved, i.e.\ closed:
\begin{equation}\label{eq:conservation-law}
    \mathrm dJ^{(p)}=0\quad \forall \; J^{(p)}\in \mathcal J.
  \end{equation}
  Furthermore, we ignore those currents which are exact. Thus, the space of conserved Noether currents is
  \begin{equation}
    \label{eq:conservednoethercurrents}
    \mathcal J_\mathrm{Noether}= \ker(\mathrm d)\subset \Gamma_M\mleft(\bigoplus_{p=0}^{d-1}\mathrm T^{*\wedge p}M\otimes E^{(p)}\mright)
  \end{equation}
When the bundles \(E^{(p)}\) are not trivial, in general the definition of the operator \(\mathrm d\) requires the choice of a connection on \(E^{(p)}\).
We therefore pick a connection on each \(E^{(p)}\).

We now consider an action principle whose equations of motion include \eqref{eq:conservation-law}:
\begin{equation}\label{eq:BF-action-body}
    S = \int_M \sum_{p=0}^{d-1}\Lambda^{(d-p-1)}\wedge\mathrm dJ^{(p)},
\end{equation}
where, for \(p\ne0\),
\begin{equation}\Lambda^{(d-p-1)}\in\Omega^{d-p-1}(M;E^{(p)*}\otimes\operatorname{or}_M)\end{equation}
is a Lagrange multiplier \((d-p-1)\)-form field valued in the dual bundle of \(E^{(p)}\) twisted by \(\operatorname{or}_M\) that enforces \eqref{eq:conservation-law}. We can immediately recognise this as being the action of a \(BF\) theory.

For \(p=0\), the situation is slightly more subtle: the Lagrange multiplier \(\Lambda^{(d-1)}\) is valued in the dual of the vertical bundle\footnote{The vertical bundle \cite[§10.1.1]{Nakahara:2003nw} of a smooth fibre bundle \(\pi\colon Y\to X\) is the subbundle of \(\mathrm TY\) given by the kernel of \(\mathrm D\pi\colon\mathrm TY\to\mathrm TX\).} of \(\Ep{0}\twoheadrightarrow M\) (pulled back to \(M\) via \(J^{(0)}\colon M\to \Ep{0}\)), twisted by \(\operatorname{or}_M\). That is, \((J^{(0)},\Lambda^{(d-1)})\) together form a section of the bundle
\begin{equation}
    \mathrm T^{*\wedge(d-1)}M\otimes\operatorname{or}_M\otimes\mathrm V^*\Ep{0}
\end{equation}
over \(M\), whose fibre at a single point \(x\in M\) is the total space
\begin{equation}
    \mathrm T_x^{*\wedge(d-1)}M\otimes(\operatorname{or}_M)_x\otimes\mathrm T^*(\Ep{0}_x)
\end{equation}
of a particular vector bundle over \(\Ep{0}_x\).

That is, the fields \((J^{(p)},\Lambda^{(d-p-1)})_{0\le p\le d-1}\) live in the graded manifold
\begin{multline}
    \tilde X_{BF} = \Gamma_M\Bigg(\bigoplus_{p=1}^{d-1}\Big(\underbrace{\mathrm T^{*\wedge p}M\otimes E^{(p)}}_{J^{(p)}}
    \oplus
    \underbrace{\mathrm T^{*\wedge(d-p-1)}M\otimes E^{(p)*}\otimes\operatorname{or}_M}_{\Lambda^{(d-p-1)}}\Big)\\
    \times
    \underbrace{\mathrm T^{*\wedge(d-1)}M\otimes\operatorname{or}_M\otimes\mathrm V^*\Ep{0}}_{(J^{(0)},\Lambda^{(d-1)})}
    \Bigg).
\end{multline}
The graded manifold \(\tilde X_{BF}\) does not yet suffice for the Batalin--Vilkovisky formalism because we have yet to introduce antifields. The graded manifold corresponding to the Batalin--Vilkovisky description of the \(BF\) model is\footnote{Contrary to what is done in e.g.\ \cite{Jurco:2018sby}, we do not include a tower of ghosts in this graded manifold. This is because, when the connection on \(E^{(p)}\) is not flat, then it is in general not the case that \(\mathrm d\circ\mathrm d=0\), so that gauge symmetries may not exist. This is the case for the stress--energy tensor, valued in \(E^{(d-1)}=\mathrm TM\otimes\operatorname{or}_M\), on a generic manifold \(M\).}: 
\begin{multline}\label{eq:BVBF}
    X_{BF} = \Gamma_M\Bigg(\bigoplus_{p=1}^{d-1}\Big(\underbrace{\mathrm T^{*\wedge p}M\otimes E^{(p)}}_{J^{(p)}}
    \oplus
    \underbrace{\mathrm T^{*\wedge(d-p-1)}M\otimes E^{(p)*}\otimes\operatorname{or}_M}_{\Lambda^{(d-p-1)}}\\
    \oplus
    \underbrace{\mathrm T^{*\wedge(d-p)}M\otimes E^{(p)*}[-1]\otimes\operatorname{or}_M}_{J^{(p)+}}
    \oplus
    \underbrace{\mathrm T^{\wedge(p+1)}M\otimes E^{(p)}[-1]}_{\Lambda^{(d-p-1)+}}\Big)\\
    \times
    \Big(
    \underbrace{\mathrm T^{*\wedge(d-1)}M\otimes\operatorname{or}_M\otimes\mathrm V^*\Ep{0}}_{\Lambda^{(d-1)}}
    \oplus_{\Ep{0}}
    \underbrace{\mathrm T^*M\otimes\mathrm V[-1]\Ep{0}}_{\Lambda^{(d-1)+}}\\
    \oplus_{\Ep{0}}
    \underbrace{\Ep{0}}_{J^{(0)}}\oplus_{\Ep{0}}
    \underbrace{\mathrm T^{*\wedge d}M\otimes\operatorname{or}_M\otimes\mathrm V^*[-1]\Ep{0}}_{J^{(0)+}}
    \Big)
    \Bigg).
  \end{multline}
  The antifields \((J^{(p)+},\Lambda^{(d-p-1)+})_{0\le p\le d-1}\) of the \(BF\) theory live in the (degree-shifted) spaces that naturally pair with \(\tilde X_{BF}\).

  The homological vector field on the graded manifold \(X_{BF}\) is determined by the action \eqref{eq:BF-action-body} as
\begin{equation}\label{eq:BF_hydro}
    Q_{BF} = \sum_{p=0}^{d-1}\left(\mathrm dJ^{(p)}\frac\delta{\delta\Lambda^{(d-p-1)+}}
    +
    \mathrm d\Lambda^{(d-p-1)}\frac\delta{\delta J^{(p)+}}\right),
\end{equation}
enforcing the equations of motion, which are \(\mathrm dJ^{(p)}=0\) and \(\mathrm d\Lambda^{(d-p-1)}=0\).

Of course, the currents \(J^{(p)}\) are actually expressions \(J^{(p)}(\varphi,\partial\varphi,\dotsc)\) built out of the coordinates \(\varphi\) of \(X_\mathrm{micro}\)---the original field theory with which we started---that express the microscopic degrees of freedom that comprise the currents. In general, the conservation laws for \(J^{(p)}\) only hold on shell; in the Batalin--Vilkovisky formalism, this corresponds to \(\mathrm dJ^{(p)}\) being \(Q\)-exact \cite{Borsten:2025pbx}:
\[
    \mathrm d\left(J^{(p)}(\varphi,\partial \varphi,\dotsc)\right)= Q\left(Y^{(p)}(\varphi,\partial\varphi,\dotsc)\right),
\]
which is a form of the descent equation familiar from the theory of anomalies \cite{Bertlmann:1996xk} and topological field theories \cite{Labastida:2005zz}. (If \(J^{(p)}\) happens to be conserved off shell, then \(Y^{(p)}=0\).) This is to be matched to the \(BF\) theory, where
\begin{equation}
    \mathrm dJ^{(p)}=Q\Lambda^{(d-p-1)+}.
\end{equation}
This gives rise to a morphism of differential graded manifolds
\begin{equation}\label{eq:micro-to-BF}
\begin{aligned}
    X_\mathrm{micro} &\to X_{BF}\\
    \varphi &\mapsto \Big(J^{(p)}(\varphi,\partial\varphi,\dotsc),\underbrace{0,0}_{\mathclap{\Lambda^{(d-p-1)},J^{(p)+}}},\underbrace{Y(\varphi,\partial\varphi,\dotsc)}_{\Lambda^{(d-p-1)+}}\Big).
\end{aligned}
\end{equation}
This preserves the homological vector field because the action of \(Q\) on the \(BF\)-theory fields continues to be the same once the \(BF\)-theory fields are expressed in terms of those of the microscopic theory, and so is indeed a morphism of differential graded manifolds.
The map \eqref{eq:micro-to-BF} is however not a symplectomorphism but only a morphism of differential graded manifolds; the above morphism assigns zero to the Lagrange multipliers \(\Lambda^{(d-p-1)}\) or the antifields \(J^{(p)+}\) and so cannot hope to preserve the symplectic form under pullback.

\subsection{The map from the hydrodynamic theory to the \texorpdfstring{\(BF\)}{𝐵𝐹} theory}
For hydrodynamics, we will deviate slightly from the Batalin--Vilkovisky formalism described in \cref{ssec:BV} because hydrodynamic equations do not always naturally derive from an action principle due to dissipation.\footnote{
    However, see e.g.\ \cite{1988AnRFM..20..225S,RevModPhys.70.467,PhysRevE.59.5482,gomes2005variational,Fukagawa_2012,10.1063/5.0175959} for a discussion of action principles for hydrodynamics.
}
Instead, we will describe the partial differential equations of hydrodynamics in terms of a differential graded manifold \((X_\mathrm{hydro},Q_\mathrm{hydro})\) without a symplectic structure; the equations of motion are encoded in the homological vector field \(Q_\mathrm{hydro}\), but the lack of a symplectic structure means that there is no action principle.

The differential graded manifold \(X_\mathrm{hydro}\) is parameterised by fields and antifields; since \(X_\mathrm{hydro}\) does not come equipped with a symplectic structure, there will not be a one-to-one correspondence between fields and antifields.

Let the hydrodynamic variables \(v\in\Gamma_M(E_\mathrm{hydro})\) take values in a graded fibre bundle \(E_\mathrm{hydro}\), so that the space of fields is \(\Gamma_M(E_\mathrm{hydro})\). 
The antifields, on the other hand, are in one-to-one correspondence with the equations of motion.\footnote{This holds even when a symplectic structure is absent. When a symplectic structure, and hence an action, is present, then the Euler--Lagrange variational equations are such that every field corresponds to a corresponding variational equation of motion (i.e.\ an antifield), and this pairing is the symplectic pairing.}
The hydrodynamic theory describes the conservation of the currents \(J^{(p)}\) discussed in \cref{ssec:BF-model}, so that the equations of motion for the hydrodynamic theory are in bijection with (those equations of motion corresponding to) the antifields \(\Lambda^{(d-p-1)+}\) in \(X_{BF}\). By abuse of notation, we use the same symbol for both the antifields of \(X_{BF}\) and the antifields of \(X_\mathrm{hydro}\).

The antifield \(\Lambda^{(d-1)+}\) requires special treatment since by itself it is not a section of a bundle; rather, it lives on a bundle over \(E^{(0)}\), in which the zero-form current \(J^{(0)}\) takes values.
In hydrodynamics, \(J^{(0)}\) is to be expressed in terms of (derivatives of) hydrodynamic variables; this corresponds to a bundle map
\begin{equation}
    \mathrm J^\infty E_\mathrm{hydro} \to \Ep{0},
\end{equation}
where \(\mathrm J^\infty E_\mathrm{hydro} = \bigcup_{k=0}^\infty \mathrm J^kE_\mathrm{hydro}\) is the infinite-order jet bundle (see e.g.\ \cite{Sardanashvily:2009br}) for \(E_\mathrm{hydro}\), whose fibre parameterises all derivatives of sections of \(E_\mathrm{hydro}\). On the other hand, we have the canonical projection map
\begin{equation}
    \mathrm T^*M\otimes\mathrm V[-1]\Ep{0} \twoheadrightarrow \Ep{0},
\end{equation}
whose fibres parameterise \(\Lambda^{(d-1)+}\). Admissible pairs of \((v,\Lambda^{(d-1)+})\) are such that, under the projection maps
\begin{equation}
\begin{aligned}
    \pi_1&\colon&\Gamma(E_\mathrm{hydro}) \xrightarrow{\text{\shortstack{jet pro-\\longation}}} \Gamma(\mathrm J^\infty E_\mathrm{hydro}) &\to \Gamma(\Ep{0}) \\
    \pi_2&\colon&\Gamma(\mathrm T^*M\otimes\mathrm V[-1]\Ep{0}) &\twoheadrightarrow \Gamma(\Ep{0}),
\end{aligned}
\end{equation}
both \(\pi_1(v)\) and \(\pi_2(\Lambda^{(d-1)+})\) must map to the same zero-form current \(J^{(0)}\). The set of all such admissible pairs is called a fibred product (or pullback, in category-theoretic parlance), notated as \(\mathbin{{}_{\pi_1}\!\times_{\pi_2}}\). Therefore, we describe the hydrodynamic theory by means of the differential graded manifold
\begin{equation}
    X_\mathrm{hydro} = 
    \underbrace{\Gamma_M(E_\mathrm{hydro})\mathbin{{}_{\pi_1}\!\times_{\pi_2}}\Gamma(\mathrm T^*M\otimes\mathrm V[-1]\Ep{0})}_{(v,\Lambda^{(d-1)+})}\times
    \Gamma_M\Bigg(
    \bigoplus_{p=1}^{d-1}
    \underbrace{
    \mathrm T^{*\wedge(p+1)}M\otimes E^{(p)}
    }_{\Lambda^{(d-p-1)+}}
    \Bigg)
\end{equation}
with differential
\begin{equation}\label{eq:Q_hydro-general}
    Q_\mathrm{hydro}=\sum_{p=0}^{d-1}\mathrm dJ^{(p)}(v,\partial v,\dotsc)\frac\delta{\delta\Lambda^{(d-p-1)+}}.
\end{equation}
The assumption in hydrodynamics is that the Noether currents \(J^{(p)}\) can be expressed in terms of the hydrodynamic variables \(v\). This defines a morphism of differential graded manifolds
\begin{equation}\label{eq:hydro-to-BF}
    X_\mathrm{hydro}\to X_{BF}
\end{equation}
that maps the antifields \((\Lambda^{(d-p-1)+})_{0\le p\le d-1}\) in \(X_\mathrm{hydro}\) (in bijection with the conservation equations) to the corresponding antifields in \(X_{BF}\) while mapping the hydrodynamic variables \(v\) to the hydrodynamic currents \(J^{(p)}(v)\) parameterised in terms of the hydrodynamic variables; note that it preserves the homological vector field because \eqref{eq:Q_hydro-general} has the same form as the first term of \eqref{eq:BF_hydro} (and the second term of \eqref{eq:BF_hydro} only concerns \(J^+\) and \(\Lambda\) that do not exist in \(X_\mathrm{hydro}\)).
This morphism \eqref{eq:hydro-to-BF}, combined with \eqref{eq:micro-to-BF}, forms a cospan of differential graded manifolds
\begin{equation}
    \begin{tikzcd}[row sep = 1 ex, ampersand replacement=\&]
      \& X_{BF} \&  \\
            X_\mathrm{micro} \ar{ur} \& \&\ar{ul} X_\mathrm{hydro}
    \end{tikzcd}
\end{equation}
that describe the two different ways of obtaining the conserved currents (either in terms of the microscopic degrees of freedom or in terms of the hydrodynamic variables).

We note that the differential graded manifold describing a system of partial differential equations is invariant with respect to changes of coordinates; when one linearises at a specific point, the change of local coordinates induces a quasi-isomorphism of \(L_\infty\)-algebras associated to that point. In particular, hydrodynamic equations admit a description either in terms of Eulerian or Lagrangian variables, and this corresponds to a diffeomorphism of differential graded manifolds or (when linearised) a quasi-isomorphism of \(L_\infty\)-algebras describing the hydrodynamic equations.

\subsection{Overdetermined systems in hydrodynamics with higher\texorpdfstring-{‐}form symmetries}
For \(p\)-form currents with sufficiently low \(p\), it can happen that the currents \(J^{(p)}\) have \emph{fewer} components than the number of equations. In this case, one obtains an \emph{overdetermined} system of equations.

For example, consider hydrodynamics describing solely the magnetic \((d-3)\)-form symmetry of Maxwell theory with current the field strength \(F\) (whose closedness is the Bianchi identity). The conservation equation \(\mathrm dF=0\) comprises \(\binom d3\) independent equations for \(\binom d2\) components of \(F\). In \(d\le5\) this is not overdetermined, but for \(d\ge6\) this is overdetermined.

In this case, if \(J^{(p)}\) takes values in a flat bundle \(E^{(p)}\), instead of taking a hydrodynamic ansatz, one can simply postulate the potential \(A^{(p-1)}\) associated to the current \(J^{(p)}\) such that \(\mathrm dA^{(p-1)}=J^{(p)}\) solves the conservation equation. This is a generalisation of, for instance, the situation in static irrotational flow.
For a static irrotational fluid in \(d\) spatial dimensions (we ignore time for simplicity), the velocity vector field \(\vec u\) defines one-form current
\begin{equation}
    u^\flat \coloneqq u_i\,\mathrm dx^i,
\end{equation}
such that the conservation equation
\begin{equation}\label{eq:irrotationality-constraint}
    \mathrm du^\flat=0
\end{equation}
is equivalent to the irrotationality constraint.
The conservation equation \(\mathrm du^\flat=0\) then has \(\binom d2\) components, which can exceed the number of components of \(u^\flat\) (which is \(d\)); in this case it is helpful to employ the velociy potential \(\phi\) such that \(u^\flat=\mathrm d\phi\) automatically solves (locally) the conservation equation \eqref{eq:irrotationality-constraint}.

\section{Examples}\label{sec:examples}
\subsection{Relativistic hydrodynamics}\label{ssec:relativistic-hydrodynamics}
For the microscopic theory, for simplicity, consider a real scalar field \(\phi\) with self-interaction governed by a potential \(V(\phi)\):
\begin{equation}\label{eq:scalar-action}
    S = \int \mathrm d^dx\,\sqrt{|\det g|}\left(-\frac12g^{\mu\nu}(\partial_\mu\phi)(\partial_\nu\phi)-V(\phi)\right).
\end{equation}
Again for simplicity, we will describe this system using relativistic Euler's equations. The microscopic theory is given by the graded manifold
\begin{equation}
    X_\mathrm{micro} = \underbrace{\mathcal C^\infty(M)}_{\phi}\times\underbrace{\Omega^d(M;\operatorname{or}_M)[-1]}_{\phi^+},
\end{equation}
where \(\mathcal C^\infty(M)\) is the space of smooth real-valued functions on \(M\) and \(\Gamma_M(\mathrm T^{*\wedge d}M\otimes\operatorname{or}_M)\) is the space of densities on \(M\).
The symplectic pairing \(\omega\) then pairs the fields and antifields according to
\begin{equation}\label{eq:scalar-symplectic}
    \omega_\mathrm{micro}(\phi,\phi^+) = \int_M \phi\phi^+,
\end{equation}
which is well defined if at least one of \(\phi\) or \(\phi^+\) is compactly supported.
The homological vector field then encodes the equations of motion:
\begin{equation}
    Q_\mathrm{micro} = \left(\laplac\phi(x)-V'(\phi(x))\right)\frac\delta{\delta \phi^+(x)}.
\end{equation}

The conserved quantity is the stress--energy tensor, which we can dualise into a vector-valued \((d-1)\)-form according to \eqref{eq:dualise}, so that the \(BF\) theory is described by the graded manifold
\begin{multline}\label{eq:BF-hydrodynamics}
    X_{BF} = \overbrace{\Omega^{d-1}(M;\mathrm TM\otimes\operatorname{or}_M)}^{T^\nu} \times \overbrace{\Omega^0(M;\mathrm T^*M)}^{\Lambda_\nu}\\
    \times
    \underbrace{\Omega^1(M;\mathrm T^*M)[-1]}_{T^+_\nu} \times \underbrace{\Omega^d(M;\mathrm TM\otimes\operatorname{or}_M)[-1]}_{\Lambda^{+\nu}}
\end{multline}
with coordinates given by \(T^\nu\in\Omega^{d-1}(M;\mathrm TM\otimes\operatorname{or}_M)\) and \(\Lambda_\nu\in\Omega^0(M;\mathrm T^*M)\)
and the corresponding antifields \(T^+_\nu\in\Omega^1(M;\mathrm T^*M)[-1]\) and \(\Lambda^{+\nu}\in\Omega^d(M;\mathrm TM\otimes\operatorname{or}_M)[-1]\). The homological vector field of $X_{BF}$ encodes the conservation equation for \(T^\nu\) and a corresponding equation of motion for \(\Lambda_\nu\):
\begin{equation}\label{eq:Q-BF-hydrodynamics}
    Q_{BF} = \mathrm dT^\nu\frac\delta{\delta\Lambda^{+\nu}}
    + \mathrm d\Lambda_\nu\frac\delta{\delta T^+_\nu},
\end{equation}
corresponding to equations of motion for the \(BF\) action
\begin{equation}
    S_{BF} = \int_M\Lambda_\nu\wedge\mathrm dT^\nu.
\end{equation}

Finally, for the hydrodynamic theory, the graded manifold is
\begin{equation}\label{eq:X_hydro}
    X_\mathrm{hydro} = \underbrace{\mathcal C^\infty(M)}_\rho \times \underbrace{\{u\in\Gamma_M(\mathrm TM)|u^\mu u_\mu = -1\}}_{u^\mu} \times
    \underbrace{\Omega^d(M;\mathrm TM\otimes\operatorname{or}_M)[-1]}_{\Lambda^{+\nu}}
\end{equation}
with the homological vector field encoding the Euler equations:
\begin{equation}\label{eq:Q_hydro}
    Q_\mathrm{hydro} = \left(\nabla_\mu (\rho+p(\rho))u^\mu u^\nu + p(\rho)g^{\mu\nu}\right)\frac\delta{\delta\Lambda^{+\nu}},
\end{equation}
where \(p(\rho)\) is the equation of state describing the pressure in terms of the energy density \(\rho\). There is no symplectic structure and hence no action.

The differential-graded-manifold morphism \(X_\mathrm{micro} \to X_{BF}\) is given in coordinates by the map
\begin{equation}
    (\phi,\phi^+)\mapsto \Big(\underbrace{\partial^\nu\phi\star\mathrm d\phi - \left(\tfrac12(\partial\mu)^2 + V(\phi)\right)\star\mathrm dx^\nu}_{T^\nu},\underbrace{0,0}_{\Lambda_\nu,T^+_\nu},\underbrace{\phi^+\partial^\nu\phi}_{\Lambda^{+\nu}}\Big).
\end{equation}
The only nonzero component are (the Hodge dual of) the stress--energy tensor
\begin{equation}\label{eq:scalar-stress-energy}
    T_{\mu\nu}
    =\partial_\mu\phi\partial_\nu\phi - g_{\mu\nu}\left(\frac12(\partial\mu)^2 + V(\phi)\right)
\end{equation}
of the scalar field theory action \eqref{eq:scalar-action}, as well as the expression for \(\Lambda^{+\nu}\), which appears since the conservation law for the stress--energy tensor
\begin{equation}
    \partial_\mu T^{\mu\nu}=\left(\laplac\phi-V'(\phi)\right)\partial^\nu\phi
\end{equation}
only holds up to the equation of motion \(\laplac\phi-V'(\phi)=0\) corresponding to the antifield \(\phi^+\).

The differential-graded-manifold morphism \(X_\mathrm{hydro}\to X_{BF}\) is given in coordinates by the map
\begin{equation}
    (\rho,u,\Lambda^{+\nu})\mapsto
    \Big(\underbrace{(\rho+p(\rho))u^\nu\star u^\flat + p(\rho)\star\mathrm dx^\nu}_{T^\nu},\underbrace{0,0}_{\Lambda_\nu,T^+_\nu},\Lambda^{+\nu}\Big),
\end{equation}
where the first component corresponds to the perfect-fluid stress--energy tensor \eqref{eq:perfect-fluid}.

\subsection{Relativistic irrotational fluid}
Consider the previous example of a scalar field in \cref{ssec:relativistic-hydrodynamics}, except now that the scalar field \(\phi\colon M\to\mathbb R/(C\mathbb Z)\) is valued in the circle \(\mathbb R/(C\mathbb Z)\) of circumference \(C\). The action is still formally the same as \eqref{eq:scalar-action}, but now we require that the potential \(V(\phi)\) appearing therein be periodic:
\begin{equation}V(\phi)=V(\phi+C).\end{equation}
In addition to a stress--energy tensor \eqref{eq:scalar-stress-energy}, the theory now also enjoys a closed one-form current:
\begin{equation}\label{eq:one-form-current}
    J = \mathrm d\phi.
\end{equation}
Note that \(J\) is not exact since \(\phi\) takes values in a circle.
We can describe this using an `irrotational relativistic hydrodynamics' where, in addition to \(\nabla_\mu T^{\mu\nu}=0\), we also have \(\nabla_{[\mu}J_{\nu]}=0\), where the latter may be seen as a relativistic analogue of the condition of irrotationality \(\partial_{[i}u_{j]}=0\) in nonrelativistic fluids.

The microscopic theory is given by the graded manifold
\begin{equation}
    X_\mathrm{micro} = \underbrace{\mathcal C^\infty(M,\mathbb R/C\mathbb Z)}_\phi\times\underbrace{\Omega^d(M;\operatorname{or}_M)[-1]}_{\phi^+},
\end{equation}
where \(\mathcal C^\infty(M,\mathbb R/C\mathbb Z)\) is the space of smooth circled-valued functions on \(M\); note that the antifields do \emph{not} see the periodicity of \(\phi\) since they live in the cotangent bundle of the space of fields.
The homological vector field then encodes the equations of motion:
\begin{equation}
    Q_\mathrm{micro} = \left(\laplac\phi-V'(\phi)\right)\frac\delta{\delta \phi^+}.
\end{equation}

The conserved quantities are the dualised stress--energy tensor (a vector-valued \((d-1)\)-form) and a one-form, so that the \(BF\) theory is described by the graded manifold
\begin{equation}
    X'_{BF} = X_{BF} \times \underbrace{\Omega^1(M)}_J \times \underbrace{\Omega^{d-2}(M;\operatorname{or}_M)}_\lambda
    \times
    \underbrace{\Omega^{d-1}(M;\operatorname{or}_M)[-1]}_{J^+} \times \underbrace{\Omega^2(M)[-1]}_{\lambda^+},
\end{equation}
where \(X_{BF}\) is that given in \eqref{eq:BF-hydrodynamics}, and \(X'_{BF}\) has additional factors corresponding to the other conserved current \(J\) and the corresponding Lagrange multiplier \(\lambda\). The homological vector field encodes the conservation equations for \(T^\nu\) and \(J\) and corresponding equations of motion for \(\Lambda_\nu\) and \(\lambda\):
\begin{equation}
    Q'_{BF} = Q_{BF} + \mathrm dJ\frac\delta{\delta\lambda^+}
    + \mathrm d\lambda\frac\delta{\delta J^+},
\end{equation}
where \(Q_{BF}\) is that given in \eqref{eq:Q-BF-hydrodynamics},
corresponding to equations of motion for the \(BF\) action
\begin{equation}
    S = \int_M\Lambda_\nu\wedge\mathrm dT^\nu+\lambda\wedge\mathrm dJ.
\end{equation}
For the hydrodynamic theory, the graded manifold is
\begin{equation}
    X'_\mathrm{hydro} = X_\mathrm{hydro} \times \underbrace{\Omega^2(M)[-1]}_{\lambda^+},
\end{equation}
where \(X_\mathrm{hydro}\) was given in \eqref{eq:X_hydro}. The homological vector field describes the Euler equations as well as the irrotationality constraint \(\mathrm dJ=0\):
\begin{equation}
    Q'_\mathrm{hydro} = Q_\mathrm{hydro} + \left(\nabla_\mu(q(\rho)u^\mu)\right)\frac\delta{\delta\lambda^+},
\end{equation}
where \(Q_\mathrm{hydro}\) was given in \eqref{eq:Q_hydro} and we have parameterised
\begin{equation}\label{eq:J-parameterisation}
    J = q(\rho)u_\mu\,\mathrm dx^\mu = q(\rho)u^\flat
\end{equation}
for some function \(q\) of \(\rho\).\footnote{One could choose other parameterisations, possibly including new parameters; we have chosen this parameterisation for simplicity.}

For sufficiently high number of spacetime dimensions \(d\), this is an overdetermined system since the number of equations is \(d+\binom d2\) while the number of unknowns is \(d\).

The differential-graded-manifold morphism \(X'_\mathrm{micro} \to X'_{BF}\) is given in coordinates by the map
\begin{equation}
    (\phi,\phi^+)\mapsto \Big(\underbrace{\partial^\nu\phi\star\mathrm d\phi - \left(\tfrac12(\partial\mu)^2 + V(\phi)\right)\star\mathrm dx^\nu}_{T^\nu}, \underbrace{0,0}_{\mathclap{\Lambda_\nu,T^+_\nu}},\underbrace{\phi^+\partial^\nu\phi}_{\Lambda^{+\nu}},\underbrace{\mathrm d\phi}_J,\underbrace{0,0,0}_{\mathclap{\lambda,J^+,\lambda^+}}\Big)
\end{equation}
using \eqref{eq:scalar-stress-energy} and \eqref{eq:one-form-current}. (Note that \(\lambda^+=0\) since \(\mathrm d\phi\) is closed off shell whereas \(\Lambda^{+\nu}\ne0\) since \(T^\nu\) is closed only on shell.)
The differential-graded-manifold morphism \(X'_\mathrm{hydro}\to X'_{BF}\) is given in coordinates by the map
\begin{multline}
    (\rho,u^\mu,\Lambda^{+\nu},\lambda^+)\mapsto\\
    \Big(\underbrace{(\rho+p(\rho))u^\nu\star u^\flat + p(\rho)\star\mathrm dx^\nu}_{T^\nu},\underbrace{0,0}_{\mathclap{\Lambda_\nu,T^+_\nu}},\Lambda^{+\nu},\underbrace{q(\rho)u^\flat}_J,\underbrace{0,0}_{\mathclap{\lambda,J^+}},\lambda^+\Big),
\end{multline}
whose first and fifth components correspond to \eqref{eq:perfect-fluid} and \eqref{eq:J-parameterisation} respectively.

\subsection[Magnetohydrodynamics of a \texorpdfstring{\(p\)-}{𝑝‐}form gauge field]{Magnetohydrodynamics of a \(\boldsymbol p\)-form gauge field}
It has been recently recognised that magnetohydrodynamics can be reformulated as the consequence of the conservation of stress--energy \(\nabla_\mu T^{\mu\nu}=0\) as well as the Bianchi identity \(\mathrm dF=0\) of the Maxwell field strength \(F\) \cite{Grozdanov:2016tdf} (see the review \cite{Iqbal:2024pee}), where the latter is an example of a \((d-3)\)-form symmetry; this then naturally admits a higher-form generalisation for the magnetohydrodynamics of a \((p-1)\)-brane fluid coupled to a \(p\)-form gauge field \cite{Armas:2018ibg}. In this section, we explain how the higher-form-symmetry conception of magnetohydrodynamics fits into our formalism.

As the microscopic degrees of freedom, let us consider a \(p\)-form gauge field \(A\) with a \((p+1)\)-form field strength \(F=\mathrm dA\) coupled to \((p-1)\)-branes. Since we are working with a second-quantised action,
for simplicity we regard the \((p-1)\)-branes as arising from a scalar field \(\phi\) defined on the \((p-1)\)st iterated loop space \(L^{p-1}M\) of spacetime, similar to the discussion in \cite{Borsten:2025diy}.\footnote{This is regarded strictly as a toy model; there are many complications with second-quantised actions of branes, which we ignore.} For this example only, to avoid complications with orientations on iterated loop spaces, we suppose that some compatible orientations on \(M\) and \(L^{p-1}M\) have been chosen.
Then the action is as follows:
\begin{equation}
    S = \int_M \mathrm dA\wedge\star\mathrm dA
    + \int_{L^{p-1}M} (\mathrm d+\check A)\phi\wedge\star(\mathrm d+\check A),
\end{equation}
where \(\check A\) is a one-form gauge field defined on \(L^{p-1}M\) by iterating the Chen-form construction \cite{zbMATH03361026,zbMATH03532253,zbMATH04050538} as explained in \cite{Kim:2019owc,Borsten:2025diy}.
This yields the stress--energy tensor
\begin{multline}\label{eq:stress-energy-brane}
T_{\mu\nu}(x)
= \frac1{p!}
F_{\mu\alpha_1\cdots\alpha_p}
F_\nu{}^{\alpha_1\cdots\alpha_p}
-
\frac1{2(p+1)!}
g_{\mu\nu}
F_{\alpha_1\cdots\alpha_{p+1}}
F^{\alpha_1\cdots\alpha_{p+1}}\\
+ 
\int_{L^{p-1}M}\mathrm D\sigma
\int_{\mathbb T^{p-1}}\mathrm d^{p-1}t\,
\delta^{(d)}(x,\sigma(t))
\left(2D_{(\mu}\phi D_{\nu)}\phi^*
-g_{\mu\nu}g^{\rho\sigma}(D_\rho\phi D_\sigma\phi^*)\right)|_\sigma,
\end{multline}
where \(\mathbb T^{p-1}\) is the \((p-1)\)-dimensional torus (the domain of elements of the iterated loop space \(L^{p-1}M\)).
Thus, the differential graded manifold describing the microscopic theory is
\begin{equation}
    X''_\mathrm{micro}
    = \overbrace{\mathrm T^*[-1]}^{\mathclap{\text{antifields}}}\left(
        \underbrace{\Omega^p(M)}_A
        \times\underbrace{\Omega^{p-1}(M)[1]
        \times\dotsb\times \Omega^0(M)[p]}_{\text{ghosts for \(p\)-form gauge field}}
        \times
        \mathcal C^\infty(L^{p-1}M)
    \right),
\end{equation}
where we have ignored the sum over gerbes (and the fact that \(p\)-form potentials need not be globally defined) for simplicity, and where we have included the tower of higher-order ghost fields that arise in the Batalin--Vilkovisky formalism \cite{Jurco:2018sby}. Due to the proliferation of ghost fields, we omit the explicit description of the homological vector field \(Q''_\mathrm{micro}\), which can be read off from the equations of motion.

The conserved quantities of interest in the microscopic theory are the stress--energy tensor \(T^\nu\) as before as well as the \((p+1)\)-form  field strength \(F=\mathrm dA\), where the conservation law is the Bianchi identity \(\mathrm dF=0\). In the \(BF\) theory, we regard the conserved quantities \(T^\nu\) and \(F\) as fundamental fields together with corresponding Lagrange multipliers.
Thus, the \(BF\) theory is given by the action
\begin{equation}
    S = \int \Lambda_\nu\wedge\mathrm dT^\nu + \lambda\wedge\mathrm dF,
\end{equation}
where \(F\) is a \((p-1)\)-form fundamental field and \(\lambda\) is a \((d-p-2)\)-form Lagrange multiplier on spacetime \(M\). Hence the corresponding differential graded manifold is
\begin{equation}
    X''_{BF} = X_{BF} \times \underbrace{\Omega^{p+1}(M)}_F \times \underbrace{\Omega^{d-p-2}(M)}_\lambda
    \times
    \underbrace{\Omega^{d-p-1}(M)[-1]}_{F^+} \times \underbrace{\Omega^{p+2}(M)[-1]}_{\lambda^+},
\end{equation}
where \(X_{BF}\) was defined in \eqref{eq:BF-hydrodynamics}, and \(X''_{BF}\) has additional factors corresponding to the other conserved current \(F\) and the corresponding Lagrange multiplier \(\lambda\). The homological vector field is then
\begin{equation}
    Q''_{BF} = Q_{BF} + \mathrm dF[-1]\frac\delta{\delta\lambda^+}
    + \mathrm d\lambda[-1]\frac\delta{\delta F^+},
\end{equation}
where \(Q_{BF}\) is that given in \eqref{eq:Q-BF-hydrodynamics}.

Finally, we now parameterise the currents \(T^\nu\) and \(F\) in terms of hydrodynamic variables. Following and generalising \cite[(2.12)]{Grozdanov:2016tdf}, we write
\begin{equation}\label{eq:F-hydro}
    F = \eta u^\flat\wedge h,
\end{equation}
where \(h\in\Omega^{d-p-2}(M)\) is a \((d-p-2)\)-form satisfying the constraints
\begin{align}
    u\intprod h &= 0, &
    \|h\|^2 = h_{\mu_1\dotso\mu_{d-p-2}}h^{\mu_1\dotso\mu^{d-p-2}}&=1,
\end{align}
and \(\eta\) is a new scalar parameter.\footnote{This parameter was called \(\rho\) in \cite[(2.12)]{Grozdanov:2016tdf}, but this conflicts with our notation.} The stress--energy tensor \(T^{\mu\nu}\) is parameterised as
\begin{multline}\label{eq:p-form-hydro-stress-energy}
    T^{\mu\nu}=T^{\mu\nu}_\mathrm{fluid}[\rho,u] - f(\eta,\eta')h^\mu{}_{\mu_2\dotso\mu_{d-p-2}}h^{\nu\mu_2\dotso\mu_{d-p-2}}\\
    -
    f'(\eta,\eta')k^\mu{}_{\mu_2\dotso\mu_{d-p-3}}k^{\nu\mu_2\dotso\mu_{d-p-3}},
\end{multline}
where \(T^{\mu\nu}_\mathrm{fluid}[\rho,u]\) is the stress--energy tensor for a perfect fluid given in \eqref{eq:perfect-fluid}, while \(\eta'\) is another scalar parameter, and \(k\in\Omega^{d-p-3}(M)\) is a \((d-p-3)\)-form\footnote{In the \(p=1\) case analysed in \cite{Grozdanov:2016tdf}, \(k=1\) is a constant zero-form that can be eliminated from the list of parameters.} such that
\begin{equation}
    \|k\|^2=k_{\mu_1\dotso\mu_{d-p-3}}k^{\mu_1\dotso\mu^{d-p-3}}=1,
\end{equation}
and \(f(\eta,\eta')\) and \(f'(\eta,\eta')\) are some more equations of state.
Then the number of independent parameters equals the number of independent equations, namely \(d+\binom d{p+2}\), for a closed system of equations.

Thus the differential graded manifold for the hydrodynamic equations is
\begin{multline}
    X''_\mathrm{hydro} = X_\mathrm{hydro} \times 
    \underbrace{\left\{
        \|h\|=1,\;u\intprod h=0\middle|
        h\in\Omega^{d-p-2}(M)
    \right\}}_h\\
    \times
    \underbrace{\left\{
        \|k\|=1\middle|
        k\in\Omega^{d-p-3}(M)
    \right\}}_k
        \times 
    \underbrace{\mathcal C^\infty(M)^2}_{\eta,\eta'}
    \times
    \underbrace{\Omega^2(M)[-1]}_{\lambda^+},
\end{multline}
where \(X_\mathrm{hydro}\) was defined in \eqref{eq:X_hydro},
with the homological vector field \(Q''_\mathrm{hydro}\) that can be read off from the hydrodynamic equations \(\nabla_\mu T^{\mu\nu}[\rho,u,h,k,\eta,\eta']=0\) and \(\mathrm dF[u,h,\eta]=0\).

As before, we therefore obtain the cospan
\begin{equation}
    X''_\mathrm{micro}\to X''_{BF}\leftarrow X''_\mathrm{hydro}.
\end{equation}
The morphisms are given by the parameterisations of the currents \eqref{eq:stress-energy-brane}, \eqref{eq:F-hydro}, and \eqref{eq:p-form-hydro-stress-energy}.

\section*{Acknowledgements}
Simon Jonsson is grateful for the hospitality of Aros Kapital, Vestagatan 6, 416 64 Göteborg, Sweden.

\newcommand\cyrillic[1]{\fontfamily{Domitian-TOsF}\selectfont \foreignlanguage{russian}{#1}}

\bibliographystyle{unsrturl}
\bibliography{biblio}
\end{document}